\documentclass[aps,prl,twocolumn,superscriptaddress,amsmath,amssymb]{revtex4-1}
\usepackage{graphicx}  % needed for figures
\usepackage{dcolumn}   % needed for some tables
\usepackage{bm}        % for math
\usepackage{verbatim}   % for math
\usepackage{graphicx}
\usepackage{mathtools}
\usepackage{hyperref}
\usepackage{multirow}
\usepackage{array}
\usepackage{float}
\usepackage{color}

\usepackage{amssymb}

\usepackage{mathrsfs}

\usepackage{dsfont}

\usepackage[caption=false]{subfig}
\usepackage{multibib}
\newcites{supp}{References}

\begin{document}
\title{Transport signatures of quasiparticle poisoning in a Majorana island}

\author{S.~M.~Albrecht}
\affiliation{Center for Quantum Devices and Station Q Copenhagen, Niels Bohr Institute, University of Copenhagen, Copenhagen 2100, Denmark}

\author{E.~B.~Hansen}
\affiliation{Center for Quantum Devices and Station Q Copenhagen, Niels Bohr Institute, University of Copenhagen, Copenhagen 2100, Denmark}

\author{A.~P.~Higginbotham}
\affiliation{Center for Quantum Devices and Station Q Copenhagen, Niels Bohr Institute, University of Copenhagen, Copenhagen 2100, Denmark}
\affiliation{JILA, University of Colorado and NIST, Boulder, Colorado 80309, USA}

\author{F.~Kuemmeth}
\affiliation{Center for Quantum Devices and Station Q Copenhagen, Niels Bohr Institute, University of Copenhagen, Copenhagen 2100, Denmark}

\author{T.~S.~Jespersen}
\affiliation{Center for Quantum Devices and Station Q Copenhagen, Niels Bohr Institute, University of Copenhagen, Copenhagen 2100, Denmark}

\author{J.~Nyg{\aa}rd}
\affiliation{Center for Quantum Devices and Station Q Copenhagen, Niels Bohr Institute, University of Copenhagen, Copenhagen 2100, Denmark}

\author{P.~Krogstrup}
\affiliation{Center for Quantum Devices and Station Q Copenhagen, Niels Bohr Institute, University of Copenhagen, Copenhagen 2100, Denmark}

\author{J.~Danon}
\affiliation{Center for Quantum Devices and Station Q Copenhagen, Niels Bohr Institute, University of Copenhagen, Copenhagen 2100, Denmark}
\affiliation{Department of Physics, NTNU, Norwegian University of Science and Technology, 7491 Trondheim, Norway}

\author{K.~Flensberg}
\affiliation{Center for Quantum Devices and Station Q Copenhagen, Niels Bohr Institute, University of Copenhagen, Copenhagen 2100, Denmark}

\author{C.~M.~Marcus}
\affiliation{Center for Quantum Devices and Station Q Copenhagen, Niels Bohr Institute, University of Copenhagen, Copenhagen 2100, Denmark}

\date{\today}

\begin{abstract}We investigate effects of quasiparticle poisoning in a Majorana island with strong tunnel coupling to normal-metal leads. In addition to the main Coulomb blockade diamonds, ``shadow'' diamonds appear, shifted by 1$e$ in gate voltage, consistent with transport through an excited (poisoned) state of the island. Comparison to a simple model yields an estimate of parity lifetime for the strongly coupled island ($\sim 1 ~ \mathrm{\mu s} $) and sets a bound for a weakly coupled island ($> 10 ~ \mathrm{\mu s} $). Fluctuations in the gate-voltage spacing of Coulomb peaks at high field, reflecting Majorana hybridization, are enhanced by the reduced lever arm at strong coupling. In energy units, fluctuations are consistent with previous measurements.\end{abstract}

\maketitle
Hybrid semiconductor-superconductor nanowire devices have been the focus of intense research in recent years \cite{Mourik:2012wk,Rokhinson:2012ep,Das:2012hi,Deng:2012gn,Churchill:2013cq,Albrecht:2016cw} primarily because they are expected to support Majorana zero modes \cite{Lutchyn:2010hp,Oreg:2010gkb}. Of particular relevance to schemes for Majorana fusion-rule testing, braiding, and Majorana-based quantum computation \cite{Aasen:2016bj,Vijay:2016ur,Landau:2016dn,vanHeck:2012bp} is the Majorana island geometry, in which the topological hybrid nanowire acquires a charging energy that lifts the degeneracy between occupied and empty Majorana states \cite{Fu:2010wy,Hutzen:2012gg,Albrecht:2016cw,vanHeck:2016ef,Sherman:2016wh}, allowing for charge readout of the state parity.

A fundamental bound to the coherence of Majorana based qubits is the parity lifetime of the Majorana state, limited by quasiparticle poisoning \cite{Rainis:2012uw,Cheng:2012bua,Aasen:2016bj}.
Studies on metallic superconductors have explored the associated poisoning rates in great detail \cite{Aumentado:2004ij,Ferguson:2006hg,Martinis:2009bd,DeVisser:2011cu,Zgirski:2011dx,Sun:2012gl,Riste:2013il,Maisi:2013cra,Pop:2014gg,vanWoerkom:2015jt},
while experiments on semiconductor-superconductor hybrids have only established bounds on the relaxation rate of quasiparticles into the subgap state \cite{Higginbotham:2015wc}, with quantitative estimates for poisoning from external sources still pending.

In this Letter, we use Coulomb blockade spectroscopy to quantify the quasiparticle poisoning time of a Majorana island.
We find the poisoning time to a state with one extra quasiparticle in the BCS continuum to be $\sim 1~ \mathrm{\mu s} $ in the regime of relatively strong coupling between the island and the leads, and bounded from below by $10~\mathrm{\mu s} $ in the less strongly coupled regime investigated in Ref.~\cite{Higginbotham:2015wc}.
Our results demonstrate transport signatures of quasiparticle poisoning in Majorana islands up to the topological phase transition and place constraints on a relevant timescale for topological quantum computation and Majorana braiding.

\begin{figure}[b]
\center
\includegraphics[width=86mm]{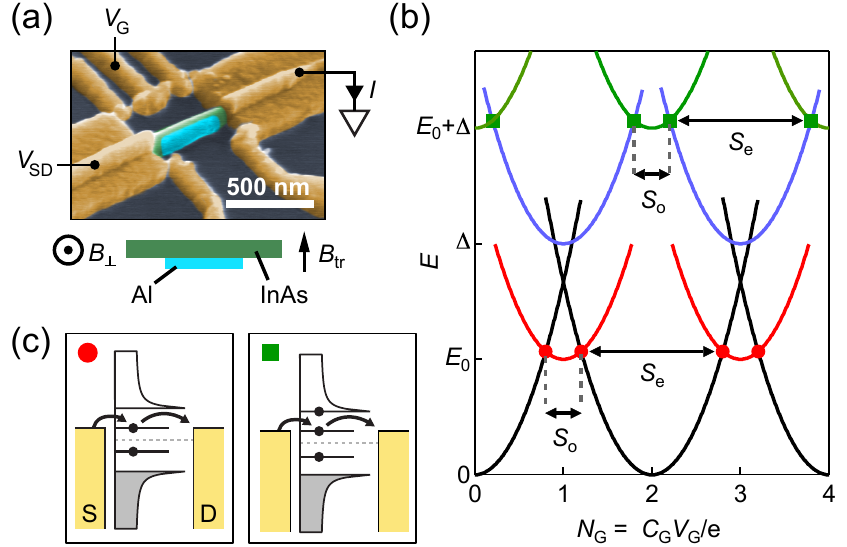}
\caption{(a) Top: Electron micrograph (false color) of a similar Majorana island device. The applied bias voltage $V_ \mathrm{SD} $, gate voltage $V_ \mathrm{G} $, and measured current $I$ are indicated. Bottom: Schematic top view (looking down onto the Si wafer) of the InAs nanowire (green) with two-facet epitaxial Al shell (light blue), showing the direction of applied magnetic fields $B_\perp$ and $B_ \mathrm{tr}$. (b) Charge-state energies of the island as a function of gate induced charge $N_\mathrm{G}$.
Spacings between degeneracies indicated with $S_\mathrm{e}$ and $S_\mathrm{o}$. (c) Schematics of transport processes for degeneracies indicated with red circles and green squares in (b).
\label{fig1}}
\end{figure}

The device we investigate consisted of an MBE-grown [0001] wurtzite InAs nanowire with epitaxial Al on two of six facets [Fig.\ \ref{fig1}(a)], which induces a hard superconducing gap in the nanowire \cite{Krogstrup:2015en,Chang:2015kwa}. The Al shell was removed on both ends using a chemical etch, leaving an Al island of length $L \sim 400~ \mathrm{nm} $.
Uncovered InAs segments at the wire ends are electrically contacted using normal-metal (Ti/Au) ohmic contacts. Lithographically patterned electrostatic gates near the $\sim50$~nm exposed segments next to the ohmic contacts were used to deplete carriers, bringing the device into the Coulomb blockade regime.
Magnetic fields were applied perpendicular to the nanowire axis, with out-of-plane field denoted $B_\perp$ and in-plane field denoted $B_\mathrm{tr}$ [Fig.\ \ref{fig1}(a), lower panel].
Due to the thin ($\sim10$~nm) Al shell on the side of the nanowire, superconductivity was preserved up to a large out-of-plane critical field, $B_{\mathrm{c},\perp} \sim 0.7$~T, and lower in-plane critical field, $B_\mathrm{c,tr} \sim 0.2$~T. The chemical potential of the island was controlled by the voltage $V_\mathrm{G}$ on a side gate. A voltage bias, $V_\mathrm{SD}$, with 5 $\mu$V AC component at 314 Hz, was used to measure differential conductance using standard lockin methods. All measurements were carried out in a dilution refrigerator with $\sim$ 50 mK base temperature. 

Before discussing experimental results, we briefly introduce a simple model of a hybrid Coulomb island with normal-metal leads (see Supplemental Material for details of the model). We take the density of states of the island to consist of a single subgap state at energy $E_0$ plus a Bardeen-Cooper-Schrieffer (BCS)-like continuum above a gap $\Delta$ \cite{Higginbotham:2015wc,Albrecht:2016cw}.
For charging energy $E_\mathrm{C}$ exceeding thermal energy, the total number of charges on the island $N$ is well defined.
We write $N = 2N_ \mathrm{cp} + N_\Delta + N_0 $, with $N_\mathrm{cp}$ the number of Cooper pairs on the island, $N_0 = 0,1$ the occupation of the subgap state, and $N_\Delta$ the number of quasiparticles in the BCS continuum.

Neglecting thermal effects, we label the available states in the Majorana island by their associated charge occupation numbers $\left(N_\mathrm{cp},N_\Delta,N_0\right)$, with corresponding charge state energies
\begin{equation}
E(N_\mathrm{cp},N_\Delta,N_0) = \frac{E_ \mathrm{C}}{2} \left(N_ \mathrm{G} - N \right)^2 + N_\Delta \Delta + N_0 E_0,
\end{equation}
where $N_\mathrm{G} =  C_ \mathrm{G} V_\mathrm{G}/e$ is the gate-induced charge on the island, $C_ \mathrm{G}$ is the capacitance between island and side gate, and all quasiparticles in the BCS-continuum are assumed to have relaxed to the gap energy, $\Delta$. The resulting spectrum is show in Fig.\ \ref{fig1}(b) as a function of $N_{\rm G}$ (modulo an even integer), where we assume $E_0 < E_{\rm C}/2 < \Delta$. For even $N$, the lowest available charge state (shown in black) is the pure condensate state $(N_ \mathrm{cp} ,0,0)$, followed by a state with an occupied subgap state and one quasiparticle in the BCS continuum $(N_ \mathrm{cp} ,1,1)$ (green).
For odd $N$, the lowest two available charge states are $(N_ \mathrm{cp} ,0,1)$ and $(N_\mathrm{cp} ,1,0)$, shown in red and blue.

We next consider transport through the island at small bias.
At low temperatures and ignoring quasiparticle poisoning for now, the island is expected to be mostly in its ground charge state, with transport occurring only at charge-state degeneracies [red circles in Fig.\ \ref{fig1}(b)].
At these points, an extra electron can be added or removed from the island without energy cost, changing the occupation of the subgap state.
Transport cycles at these degeneracies [left panel of Fig.\ \ref{fig1}(c)] correspond to the processes,
\begin{align}
\begin{split}
(N_\mathrm{cp},0,0)  &\rightleftarrows (N_\mathrm{cp},0,1),\\
\text{and }(N_\mathrm{cp},0,1) & \rightleftarrows (N_\mathrm{cp}+1,0,0).
\label{eq.main}
\end{split}
\end{align}
The two degeneracies are  symmetric about odd values of $N_\mathrm{G}$, and produce a conductance peak pattern with unequal even and odd peak spacings, $S_ \mathrm{e} = \eta^{-1}\left(E_ \mathrm{C} + 2E_0 \right)$ and $S_ \mathrm{o} = \eta^{-1}\left( E_ \mathrm{C} - 2E_0\right)$, where $\eta = e(C_\mathrm{G}/C_\mathrm{tot})$ is the gate lever arm and $C_\mathrm{tot} = e^{2}/E_\mathrm{C}$.
The peak spacing difference, $S_{\rm e} - S_{\rm o}$, is thus proportional to $E_0$ \cite{Higginbotham:2015wc} and can be used to track subgap states into the Majorana regime \cite{Albrecht:2016cw}.

Quasiparticle poisoning excites the system from its charge ground state to a state with $N_\Delta=1$ [green and blue parabolas in Fig.\ \ref{fig1}(b)].
Transport can now occur at charge-state degeneracies with $N_\Delta=1$, marked as green squares in Fig.\ \ref{fig1}(b).  The associated transport cycles [right panel of Fig.\ \ref{fig1}(c)] correspond to
\begin{align}
\begin{split}
(N_\mathrm{cp},1,0) & \rightleftarrows (N_\mathrm{cp},1,1),\\
\text{and }(N_\mathrm{cp},1,1)  &\rightleftarrows (N_\mathrm{cp}+1,1,0).
\label{eq.shadows}
\end{split}
\end{align}
Processes that bring the island back to an unpoisoned state with $N_\Delta=0$ include (i) Cooper pair recombination, $(N_{\rm cp},1,1) \to (N_{\rm cp}+1,0,0)$; (ii) quasiparticle relaxation into the subgap state, $(N_{\rm cp},1,0) \to (N_{\rm cp},0,1)$; and (iii) quasiparticle tunneling out to a lead, $(N_{\rm cp},1,N_0) \to (N_{\rm cp},0,N_0)$.
Depending on the relative magnitude of the corresponding relaxation rates, the poisoning rate, and the coupling of the subgap state to the source and drain leads $\Gamma_{\rm S,D}$, the transport cycles in Eq.\ \eqref{eq.shadows} can yield measurable conductance resonances.
As evident from Fig.\ \ref{fig1}(b), the conductance peaks in the poisoned state should occur with the same peak spacings as the unpoisoned state, $S_\mathrm{e,o} $, but shifted by 1e in gate voltage. The conductance height of the poisoned peaks contains quantitative information about the quasiparticle poisoning and relaxation rates.

Coulomb blockade spectroscopy of Majorana islands reported in Refs.\ \cite{Higginbotham:2015wc, Albrecht:2016cw} showed peaks at the unpoisoned resonances, Eq.~(\ref{eq.main}), but no features associated with the poisoned transport cycles in Eq.~\eqref{eq.shadows}.
In the present experiment we use the $400~$nm device from Ref. \cite{Albrecht:2016cw}, but lower the barriers between the island and the leads.
This increases the rates $\Gamma_{\rm S,D}$ as well as the rate of quasiparticle poisoning from the leads, giving rise to measureable transport features associated with the poisoned resonances (\ref{eq.shadows}). We have observed similar behavior in two devices and present results from one device here.

\begin{figure}
\center 
\includegraphics[width=86mm]{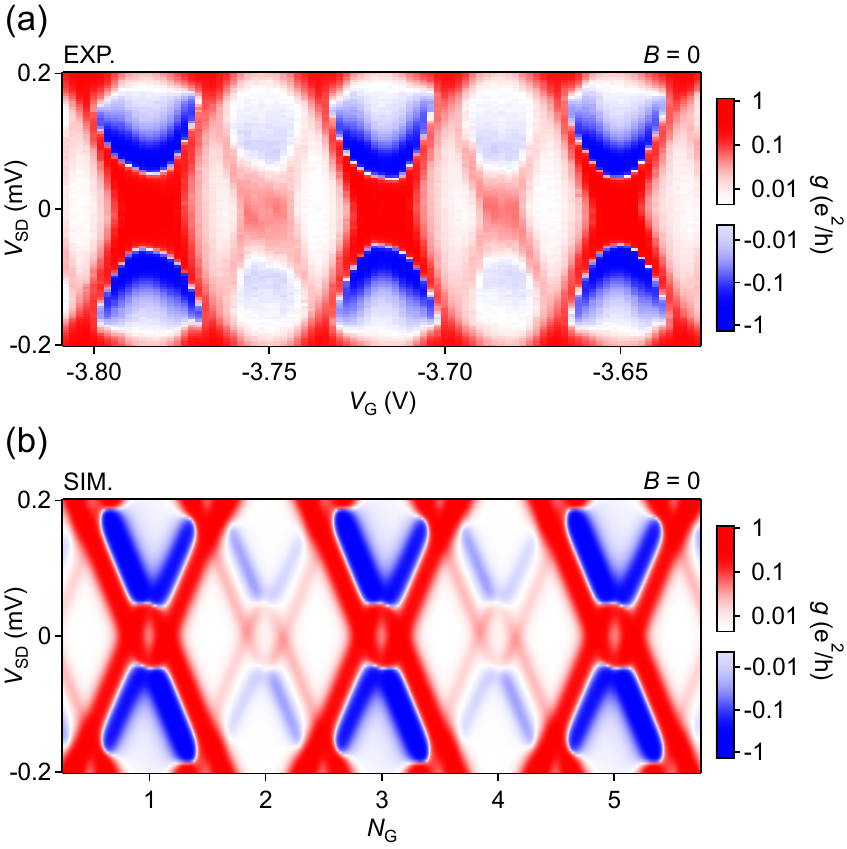}
\caption{(a) Experimental differential conductance $g$ as a function of gate voltage, $V_\mathrm{G}$, and source-drain voltage, $V_\mathrm{SD} $, at zero magnetic field, showing a series of 2$e$-periodic Coulomb diamonds with a second set of weaker ``shadow'' diamonds offset from the main diamonds by 1$e$. (b) Numerically calculated differential conductance as a function of gate-induced charge $N_\mathrm{G}$ and source-drain voltage. See main text for model parameters.\label{fig2}}
\end{figure}

Figure \ref{fig2}(a) shows the zero-bias differential conductance, $g = \mathrm{d}I/ \mathrm{d}V_\mathrm{SD}$ as a function of $V_\mathrm{SD}$ and $V_\mathrm{G}$ at zero magnetic field. The data show a high-conductance Coulomb diamond pattern with large even-occupancy diamonds, small odd-occupancy diamonds, and negative differential conductance (NDC) at finite bias, similar to previous measurements \cite{Higginbotham:2015wc,Albrecht:2016cw}. 
The nearly vanishing odd diamond indicates that the subgap state energy $E_0$ is only slightly smaller than $E_\mathrm{C}/2$ \cite{Tuominen:1992vy,Lafarge:1993zz,Higginbotham:2015wc,Albrecht:2016cw}.  

From the main Coulomb blockade diamonds we extract a charging energy $E_\mathrm{C} = 210~\mu$eV, a gate lever arm $\eta = 2E_\mathrm{C}/(\langle S_e \rangle +\langle S_o\rangle) = 6.2 $~meV/V, and a zero-field subgap state energy $E_0 = 75~\mu$eV. The width and magnitude of the conductance peak, taken at a magnetic field where peak overlap is minimal (see below), gives an asymmetric coupling of the subgap state to the source and drain leads which we fit as $\Gamma_\mathrm{S} \sim 1~\mathrm{GHz}$ and $\Gamma_\mathrm{D} \sim 6~\mathrm{GHz}$, significantly stronger than the coupling of $\sim 0.5~\mathrm{GHz}$ reported in \cite{Higginbotham:2015wc}. 

Figure~\ref{fig2}(a) shows in addition to the main Coulomb diamonds with peak conductance $g_ \mathrm{m} \sim 0.5~\mathrm{e^2}/h$, a weaker set of ``shadow'' Coulomb diamonds centered on the valleys of the main diamonds, with peak conductance $g_ \mathrm{s} \sim 0.03~\mathrm{e^2}/h$. The shadow diamonds are similar to the main diamonds, including regions of NDC, though much lower in conductance and shifted by the equivalent of 1$e$ in gate voltage. Similar shadow-like peaks were previously investigated in metallic superconductor islands \cite{Hergenrother:1994ei}, in that case made visible by increasing temperature rather than island-lead coupling.

We attribute the shadow diamonds to quasiparticle poisoning, as in Ref.~\cite{Hergenrother:1994ei}. By comparing main and shadow peak conductances to a rate-equation model of transport with poisoning, we extract a characteristic poisoning time, $\tau_\mathrm{p}$, associated with occupancy-changing excitations $(N_\mathrm{cp},0,N_0) \to ( N_ \mathrm{cp},1,N_0)$ (electronlike) and $(N_\mathrm{cp},0,N_0) \to ( N_ \mathrm{cp}-1,1,N_0)$ (holelike) in the BCS continuum.  The model does not include poisoning events that populate the subgap state as they will tunnel out again on a time scale set by the large state-lead coupling, $\Gamma_ \mathrm{D}$, and thus not contribute significantly to the shadow peak conductance. In the limit of a weakly coupled island, $\tau_{\rm p}$ sets a lower bound on the parity lifetime of the Majorana state, since in that limit only intrinsic processes based on Cooper pair breaking contribute to parity changes of the subgap state.

 Input to the model includes independently measured values for $E_ \mathrm{C} $, $\eta$, $E_0$, and $\Gamma_\mathrm{S,D}$ (see Supplementary Material). Additional parameters are the lead-continuum conductance,  $g_ \mathrm{Al}\sim 0.7~\mathrm{e^2}/h$, measured from high-bias conductance data, the induced superconducting gap, $\Delta = 140~\mu$eV, chosen to match the onset of NDC, and the relaxation time of quasiparticles from the continuum to the subgap state, previously measured to be $\tau_\mathrm{qp} = 0.1~\mu$s in similar devices \cite{Higginbotham:2015wc}. Simulated differential conductance $g$ as a function of $V_\mathrm{SD}$ and $N_\mathrm{G}$ [Fig.~\ref{fig2}(b)] reproduces the qualitative features of the experimental conductance data.
A poisoning time of $\tau_\mathrm{p} = 1.2~\mu$s gives the best agreement with the observed ratio of main and shadow-peak conductance (see below for more details).

\begin{figure}
\center
\includegraphics[width=86mm]{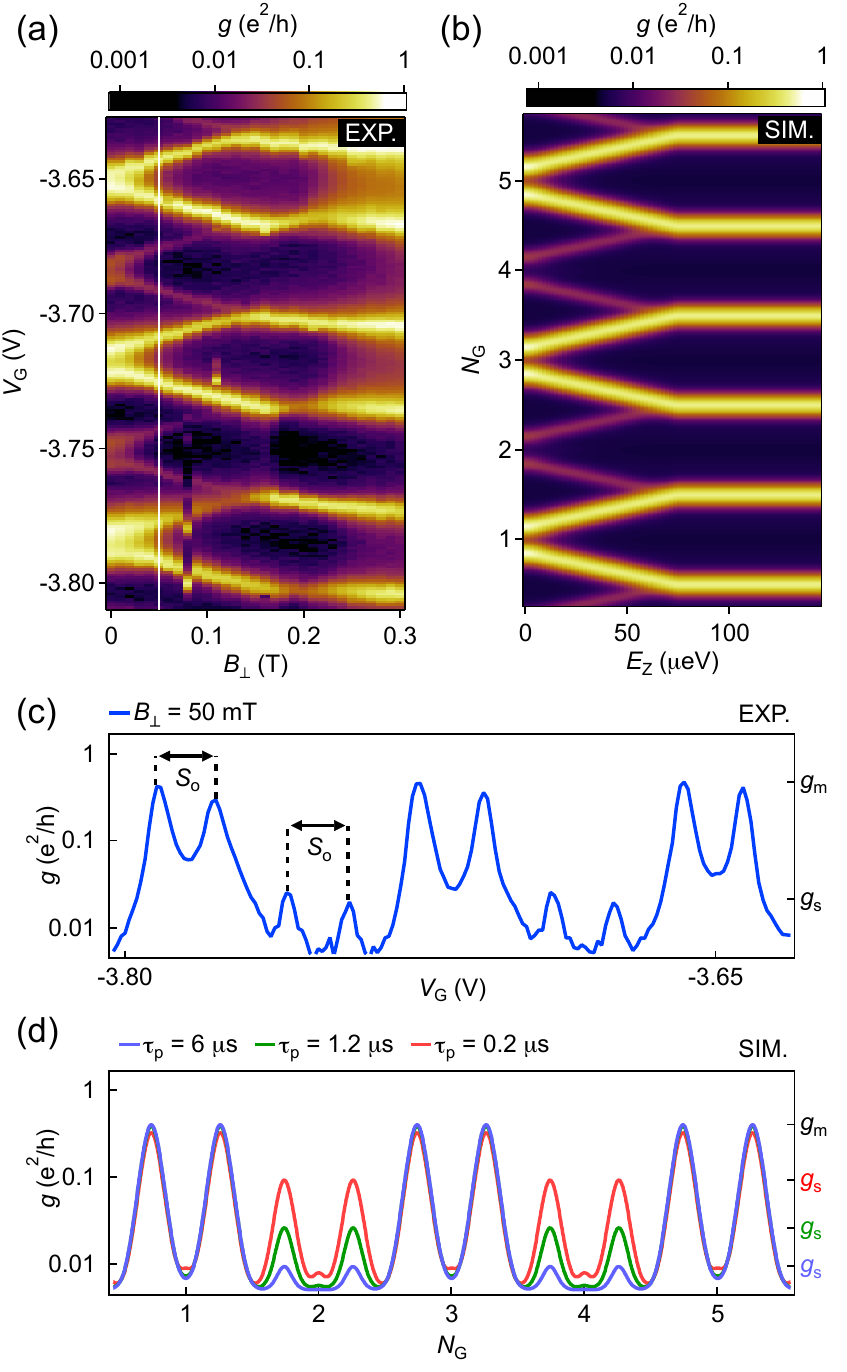}
\caption{(a) Measured zero-bias differential conductance $g$ as a function of perpendicular magnetic field, $B_\perp$, and gate voltage, $V_\mathrm{G}$, showing a series of strong even-odd Coulomb peaks with weaker shadow peaks in the even valleys between main peaks. Both sets of peaks split with increasing field and merge at $B_\perp \sim 0.16~$T.
(b) Simulated differential conductance as a function of Zeeman energy, $E_\mathrm{Z}$, and dimensionless gate voltage (charge number), $N_\mathrm{G}$.
(c) Measured differential conductance versus $V_\mathrm{G}$ at $B_\perp = 50$~mT [white line in (a)].
The average heights of the main and shadow peaks are indicated by $g_\mathrm{m}$ and $g_\mathrm{s} $ respectively.
(d) Simulated differential conductance as a function of $N_\mathrm{G}$ for poisoning times $\tau_\mathrm{p} = 0.2~\mu$s, $1.2~\mu$s, and $6~\mu$s. Simulations show an increase in $g_\mathrm{s} $ and decrease in $g_\mathrm{m} $ for decreasing $\tau_\mathrm{p}$.\label{fig3}}
\end{figure}

We now turn to the magnetic-field dependence of the transport through the island. Figure\ \ref{fig3}(a) shows the measured zero-bias differential conductance as a function of $V_\mathrm{G}$ and perpendicular magnetic field $B_\perp$.
The (initially small) odd Coulomb valley spacings $S_\mathrm{o}$ increase with $B_\perp$ up to a field of $B_\perp  \sim 0.16~$T where the average peak spacings become uniform, $\langle S_\mathrm{e} \rangle = \langle S_\mathrm{o}\rangle$, indicating a zero-energy state $E_0 = 0$.
For higher fields, the peak spacings oscillate as a function of magnetic field, as expected theoretically for hybridized Majorana modes \cite{Fu:2010wy,DasSarma:2012kt,Stanescu:2013je} and observed experimentally \cite{Albrecht:2016cw,Sherman:2016wh}.
From the near-linear dependence on $B_\perp$ of the peak spacings at lower fields  we extract an effective g-factor of 16, large for InAs \cite{Bjork:2005hx,Csonka:2008gz} but consistent with previous measurements on InAs nanowire Coulomb islands \cite{Albrecht:2016cw,Sherman:2016wh}.
Shadow peaks have the same magnetic-field dependence as the main peaks, shifted by 1$e$ gate-induced charge.
Above $B_\perp \sim 0.16~$T, where $E_0 \sim 0$, main and shadow peaks merge into one set.

The model includes the Zeeman effect by linearly lowering the subgap energy with magnetic field, $E_0 = 75~\mu{\rm eV} - E_\mathrm{Z}$, for $E_ \mathrm{Z} \leq 75~\mu$eV. To model the topological phase transition towards a Majorana mode at $E_\mathrm{Z} = 75~\mu$eV, we set $E_0 = 0$ for $E_ \mathrm{Z} > 75~\mu$eV, neglecting Majorana mode hybridization.
The resulting conductance $g$ as a function of $N_\mathrm{G}$ and $E_\mathrm{Z}$ is shown in Fig.\ \ref{fig3}(b).
Using the inferred poisoning time $\tau_\mathrm{p} = 1.2~\mu$s from Fig.\ \ref{fig2}(b) reproduces the qualitative features of the data, including the observed splitting of the main and shadow peaks for increasing $E_\mathrm{Z}$ and their merging at $E_ \mathrm{Z} = 75~\mu$eV.

A cut of the measured $g$ versus $V_\mathrm{G}$, taken at the field $B_\perp = 50~$mT (where the overlap between adjacent peaks is minimal), is shown in Fig.\ \ref{fig3}(c).
Defining $g_\mathrm{m}$ and $g_\mathrm{s}$ as the average main and shadow peak conductance, we find $g_\mathrm{m}/g_\mathrm{s} \sim 18$ in the presented gate range. Model conductance curves for three different poisoning times are show in Fig.\ \ref{fig3}(d). The model shows an increase in $g_ \mathrm{s}$ and decrease in $g_ \mathrm{m} $ for decreasing $\tau_ \mathrm{p}$. The decrease in $g_ \mathrm{m} $, de-emphasized by the logarithmic scale in Fig. \ref{fig3}(d), matches the increase in $g_ \mathrm{s} $, reflecting that the Majorana island is either in a poisoned or in an unpoisoned state. For our device-specific parameters, the model yields the simple dependence, $\tau_\mathrm{p} = a (g_\mathrm{m}/g_\mathrm{s}) + b$, with $a =0.068~\mu$s and $b =-0.004~\mu$s. From this relation and the observed ratio $g_{\rm m}/g_{\rm s}$, we infer $\tau_{\rm p} = 1.2 \pm {\rm 0.1}~\mu$s.

As the shadow peak is expected to grow with decreasing $\tau_{\rm p}$, the presented data is not taken in an optimal device tuning for the long parity lifetimes desirable for Majorana qubits. By estimating the maximum ratio $g_\mathrm{m}/g_\mathrm{s}$ from the noise-floor in more weakly-coupled device tunings, where no shadow-diamonds were observed \cite{Albrecht:2016cw}, we place a conservative estimate on the poisoning time of $\tau_\mathrm{p} > 10~\mu$s.  In the limit of a fully decoupled island this time scale of $\sim 10~\mu$s sets a conservative bound on the parity lifetime of the Majorana state.

\begin{figure}
\center
\includegraphics[width=86mm]{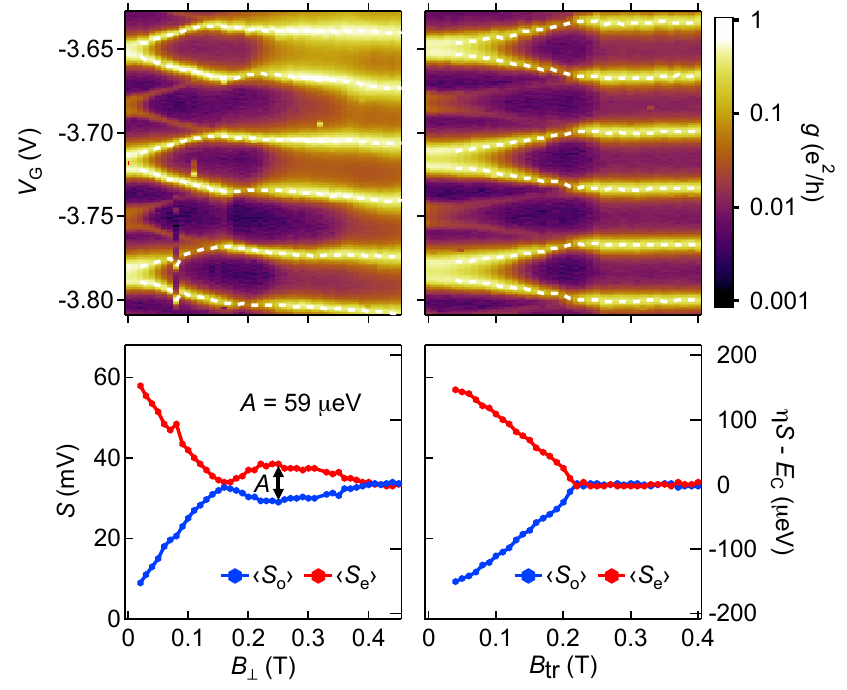}
\caption{Upper panels: Differential conductance $g$ as a function $V_\mathrm{G} $ and magnetic fields perpendicular ($B_\perp$, left) and transversal ($B_\mathrm{tr} $, right) to the nanowire.
White dashed lines indicate the fitted peak positions.
Lower panels: Extracted average peak spacing for even and odd Coulomb valleys, $\langle S_ \mathrm{e,o}\rangle$, as a function of magnetic field. The left axis has units of gate voltage, the right axis shows the associated energy scale $\eta S - E_ \mathrm{C} \propto E_0$.
\label{fig4}}
\end{figure}

Finally, we investigate the behavior of the main and shadow peaks for different magnetic field directions.
The magnetic-field dependent splitting is compared for directions $B_\perp$ and $B_ \mathrm{tr} $ in Fig. \ref{fig4}.
In both cases, the estimated peak center, indicated by a dashed white line in the upper panels of Fig. \ref{fig4}, is used to calculate average Coulomb peak spacings for the two even and the three odd valleys of the main set of Coulomb peaks, denoted $\langle S_ \mathrm{e}\rangle $ and $\langle S_ \mathrm{o} \rangle$.
The result is shown in the lower panels of Fig.\ \ref{fig4}, where the right axis indicates the energy scale for the lowest subgap state $\eta S -E_ \mathrm{C}  \propto E_0$. The shadow-peak is not used in this analysis as it cannot be distinguished from the main peak for higher fields. 

For increasing perpendicular field $B_\perp$ (left panels in Fig.\ \ref{fig4}), $\langle S_ \mathrm{e}\rangle$ and $\langle S_ \mathrm{o}\rangle $ become equal at $B_\perp \sim 0.16~$T, indicating the emergence of a state at $E_0 = 0$, and subsequently oscillate in magnetic field.
The amplitude of these oscillations, $A= 59~\mu$eV, is close to the expected value for hybridized Majorana modes in a device with $L = 400~ \mathrm{nm} $: We estimate $A = A_0 e^{-L/\xi} = 64~\mu$eV, based on previous fits of the constants $A_0 = 300~\mu$eV and $\xi = 260~\mathrm{nm}$ \cite{Albrecht:2016cw}.
We stress that data from the same device as we use here but measured in a different tuning (without shadow peaks), was included in the original analysis to determine $A_0$ and $\xi$ \cite{Albrecht:2016cw}.

For increasing in-plane fields $B_\mathrm{tr}$ (right panels in Fig.\ \ref{fig4}), the shadow peaks again split, similar to the main peaks, and acquire a 1$e$-periodic spacing at $B_\mathrm{tr} = 0.22~$T, with no oscillations visible for higher fields. Independent measurements show a closing of the superconducting gap for this device at $B_ \mathrm{c,tr} = 0.25~$T, suggesting that the transition towards 1$e$-periodic peak spacings is in this case dominated by the destruction of superconductivity. This interpretation is supported by the different curvatures of $\langle S_\mathrm{e,o}\rangle $ as they approach the field where $\langle S_ \mathrm{e}\rangle = \langle S_ \mathrm{o}\rangle$: $\langle S_ \mathrm{e,o} \rangle$ bends outwards for $B_\perp$ and inwards for $B_\mathrm{tr}$. Since $\eta \langle S_ \mathrm{e} \rangle -E_ \mathrm{C} \propto E_0$ and $\eta \langle S_ \mathrm{o} \rangle -E_ \mathrm{C} \propto -E_0$, the outward bending behavior for $B_\perp$ is in line with theoretical models for subgap states approaching the topological phase transition towards a Majorana mode \cite{DasSarma:2012kt,Rainis:2013fm,Stanescu:2013je}. In contrast, the bending inward for increasing $B_ \mathrm{tr} $ is consistent with a simple picture that approximates the subgap state energy as proportional to the quadratically-closing induced superconducting gap.

In conclusion, we have measured and modeled transport signatures of quasiparticle poisoning in a Majorana island. Zero-field measurements reveal an even-odd Coulomb diamond pattern in addition to a second set of weaker shadow diamonds, associated with quasiparticle poisoning of the Majorana island. Comparison of experiment and a simple model yields a quasiparticle poisoning time of $\tau_\mathrm{p} = 1.2~\mu$s for the presented device, and Majorana-state parity lifetimes exceeding $10~\mu$s for more weakly coupled devices where shadow features are absent. High-field measurements indicate a transition to the topological phase, with extracted Majorana mode hybridization energies consistent with previous measurements.

\vspace{0.2cm}
We thank Roman Lutchyn, Leonid Glazman, and Mingtang Deng for valuable conversations. Research supported by Microsoft Project Q, the Danish National Research Foundation, the Lundbeck Foundation, the Carlsberg Foundation, and the European Commission. CMM acknowledges support from the Villum Foundation.

%% BIBLIOGRAPHY
\bibliography{lit}
\bibliographystyle{apsrev4-1}

\clearpage
\onecolumngrid
\begin{center}
\textbf{\large Supplementary Material}
\end{center}

% Fixing numbering of equations/figures: Prefix 'S' and reset counter
\setcounter{equation}{0}
\setcounter{figure}{0}
\setcounter{table}{0}
\setcounter{page}{1}
\makeatletter
\renewcommand{\theequation}{S\arabic{equation}}
\renewcommand{\thefigure}{S\arabic{figure}}

\newcommand*{\p}{\partial}
\renewcommand*{\d}{\;\text{d}}
\renewcommand*{\u}[1]{\underline{#1}}
\newcommand*{\e}{\text{e}}
\renewcommand*{\le}{\left}
\newcommand*{\ri}{\right}
\newcommand*{\twovector}[2]{\left(\begin{array}{c}
    #1\\
    #2
  \end{array}\right)}
\newcommand*{\twomatrix}[4]{\left[\begin{array}{cc}
    #1 & #2\\
    #3 & #4
  \end{array}\right]}
\newcommand{\sigmax}{\left(\begin{array}{cc}
    0 & 1\\
    1 & 0
  \end{array}\right)}
\newcommand{\sigmay}{\left(\begin{array}{cc}
    0 & -i\\
    i & 0
  \end{array}\right)}
\begin{center}
\parbox[t][4cm][s]{0.8\textwidth}{This Supplementary Material presents more details of the numerical simulations that produced Figs.~2(b) and 3(b,d) in the main text.
We introduce the model Hamiltonian we use to describe the hybrid island, discuss its spectrum, and indicate which of the states are involved in transport through the island.
Assuming the island to be tunnel coupled to normal metallic source and drain leads, we derive from this model a set of master equations describing the basic transport dynamics of the island.
We then add phenomenological poisoning and relaxation rates that describe transitions to and from an excited state with one extra quasiparticle on the island, and we briefly discuss the possible mechanisms underlying this poisoning.
Finally, we show how we include the electronic Zeeman splitting on the island, which we need to produce the field-dependent results shown in Fig.~3 of the main text.}\end{center}

\section{Numerical simulations}

We model the nanowire as a charged metallic superconducting island, which we assume for simplicity to have a BCS continuum of states and a single spin-degenerate discrete subgap level. The island is tunnel coupled to two normal leads and the current through the wire is calculated by solving a steady state Pauli master equation. Assuming that the time between tunneling events is long enough for the island to equilibrate, we take into account only incoherent sequential tunneling transitions, which are calculated using Fermi's golden rule. We will give a brief description of the model here and refer to the supplementary material of Ref.~\citesupp{Higginbotham2015} for more detailed derivations.

\subsection{\label{sec:model}Model Hamiltonian}

The Hamiltonian of the leads, the island, and the coupling between them is
\begin{equation}
  H = H_{\text{LR}} + H_{\text{D}} + H_{\text{T}}.
\end{equation}

The leads are described by the Hamiltonian
\begin{equation}
  H_{\text{LR}} = \sum_{\alpha, \nu,\sigma}(\epsilon_{\alpha\nu}-\mu_\alpha) c^\dag_{\alpha\nu\sigma}c_{\alpha\nu\sigma},
\end{equation}
where the operator $c^\dag_{\alpha\nu\sigma}$ creates an electron with energy $\epsilon_{\alpha\nu}$ in lead $\alpha \in \{L,R\}$, with orbital index $\nu$ and spin $\sigma\in \{\uparrow,\downarrow\}$. Lead $L$ acts as the source and lead $R$ as the drain, and their chemical potentials are given by $\mu_L=-\mu_R=V_{\text{sd}}/2$ where $V_{\text{sd}}$ is a symmetrically applied bias voltage. We assume that the density of states in the leads is constant in the energy range under consideration.

The Hamiltonian describing the superconducting island with $N$ electrons on it is
\begin{equation}
  H_{\text{D}} = \sum_{\substack{\sigma\\ \zeta=e,h}}\left\{\sum_n E_n\gamma_{n\sigma,\zeta}^\dag\gamma_{n\sigma,\zeta} + E_{0\sigma} \gamma^\dag_{0\sigma,\zeta}\gamma_{0\sigma,\zeta} \right\} + E_{\text{el}}(N).\label{eq.model_H}
\end{equation}
The first term in this Hamiltonian describes the energy of the quasiparticles on the island. The quasiparticle operators are
\begin{align}
  \gamma_{n\sigma,e} & = u_n d_{n\sigma}-\sigma v_n d_{-n\bar{\sigma}}^\dag e^{-i\hat{\phi}},&
  \gamma_{n\sigma,e}^\dag & = u^*_n d_{n\sigma}^\dag-\sigma v^*_n e^{i\hat{\phi}}d_{-n\bar{\sigma}},\\
  \gamma_{n\sigma,h} & = u_n d_{n\sigma}e^{i\hat{\phi}}-\sigma v_n d_{-n\bar{\sigma}}^\dag,&
  \gamma_{n\sigma,h}^\dag & = u^*_n e^{-i\hat{\phi}}d_{n\sigma}^\dag -\sigma v^*_n d_{-n\bar{\sigma}},
\end{align}
so that
\begin{align}
  \gamma_{n\sigma,h} = \gamma_{n\sigma,e}e^{i\hat{\phi}},\qquad \qquad
  d_{n\sigma} = u^*_n\gamma_{n\sigma,e} + \sigma v_{n}\gamma_{-n\bar{\sigma},h}^\dag,
\end{align}
where $d_{n\sigma}$ is the annihiliation operator of an electron on the island with spin $\sigma$ and orbital label $n$.
The quasiparticle operator $\gamma^\dag_{n\sigma,e(h)}$ creates a quasiparticle excitation on the island which is part electron ($d^\dag_{n\sigma}$) and part hole ($d_{n\sigma}$), by adding an electron (hole) to the island.
The state $-n\bar{\sigma}$ is the time-reversed partner of $n\sigma$.
The operator $e^{\pm i\hat{\phi}}$ shifts the number of Cooper pairs on the island by $\pm 1$ and thus ensures charge conservation.
For the energies of the quasiparticles and the coherence factors on the island we use
\begin{equation}
  E_{n} = \sqrt{\epsilon_n^2 + \Delta^2}\; ,\quad |u_n|^2 = \frac{1}{2}\left(1 + \frac{\epsilon_n}{E_n}\right)\; ,\quad |v_n|^2 = \frac{1}{2}\left(1 - \frac{\epsilon_n}{E_n}\right),\label{eq:E_n}
\end{equation}
where $\epsilon_n$ is the electron energy measured from the chemical potential of the island and $\Delta$ is the effective superconducting gap on the island.
The electronic annihilation operator corresponding to the subgap state is
\begin{equation}
d_{0\sigma} = u^*_0\gamma_{0\sigma,e}+\sigma v_0 \gamma^\dag_{0\bar{\sigma},h},
\end{equation}
where $u_0$ and $v_0$ are the coherence factors of the subgap state.
The energy of the subgap state is $E_{0\sigma}$. The second term $E_{\text{el}}(N)$ in Eq.~(\ref{eq.model_H}) models the electrostatic energy of the island,
\begin{equation}
  E_{\text{el}}(N) = \frac{E_{\text{C}}}{2}(N-N_{\text{G}})^2\; ,\qquad \text{where} \quad N = \sum_{n\sigma} d_{n\sigma}^\dag d_{n\sigma} + \sum_{\sigma} d_{0\sigma}^\dag d_{0\sigma}
\end{equation}
is the number of charges on the island, and $E_{\text{C}} = \text{e}^2/C_\Sigma$ is the charging energy of the Majorana island with total capacitance $C_\Sigma$.
The dimensionless gate-induced charge number $N_{\text{G}} = C_{\text{G}}V_{\text{G}}/\text{e}$ is proportional to the gate voltage $V_{\text{G}}$ and the capacitance between the island and the gate $C_{\text{G}}$.

Lastly, expressed in terms of the quasiparticle operators the tunneling Hamiltonian reads
\begin{align}
  H_{\text{T}}
  &= \sum_{\alpha,\nu,\sigma} \Bigg\{t_{0\alpha} c_{\alpha \nu\sigma}^\dag (u^*_0\gamma_{0\sigma,e} + \sigma v_{0} \gamma_{0\bar{\sigma},h}^\dag) + t_{0\alpha}^*(u_0\gamma_{0\sigma,e}^\dag + \sigma v^*_0 \gamma_{0\bar{\sigma},h})c_{\alpha \nu\sigma}\nonumber\\
  & \hspace{4em}+ \left. \sum_{n} \left[t_{\alpha} c_{\alpha \nu\sigma}^\dag (u^*_n\gamma_{n\sigma,e} + \sigma v_{n} \gamma_{-n\bar{\sigma},h}^\dag) + t_{\alpha}^*(u_n\gamma_{n\sigma,e}^\dag + \sigma v^*_n \gamma_{-n\bar{\sigma},h})c_{\alpha \nu\sigma}\right]\right\},
\end{align}
where we assume that the tunnel coupling elements $t_\alpha$ between the electronic states in lead $\alpha$ and the quasiparticle states on the island do not depend on energy.
We, however, do allow for a different coupling strength $t_{0\alpha}$ to the subgap state.

\subsection{Spectrum}

Due to its charging energy, the total number of electrons on the island $N$ is well defined.
Therefore we can label the internal states of the island as $(N,N_\Delta,N_0)$, where $N_\Delta$ is the number of above-gap quasiparticle excitations in the BCS continuum and $N_0$ is the occupancy of the subgap state. In our transport simulations we include states with $N\in \mathbb{N}$, $N_\Delta\in\{0,1\}$, and $N_0\in\{0,\uparrow,\downarrow,2\}$, and we approximate their energies by
\begin{equation}
E(N,N_\Delta,N_0) = \frac{E_ \mathrm{C}}{2} \left(N_ \mathrm{G} - N \right)^2 + N_\Delta \Delta + N_0 E_{0}.\label{eq.spectrum}
\end{equation}
In writing (\ref{eq.spectrum}) we assume that (i) if there is a BCS quasiparticle on the island, its energy can be approximated by $\Delta$ and (ii) the supgap state energy is the same for both spin directions $E_{0\sigma}=E_0$ (and thus $N_0 \in \{ 0, 1, 2\}$).
Later we will add a finite Zeeman splitting to our model, resulting in $E_{0\uparrow} \neq E_{0\downarrow}$, see Sec.~\ref{sec:zeeman} below.

\begin{figure}[ht]
  \includegraphics[width=12cm]{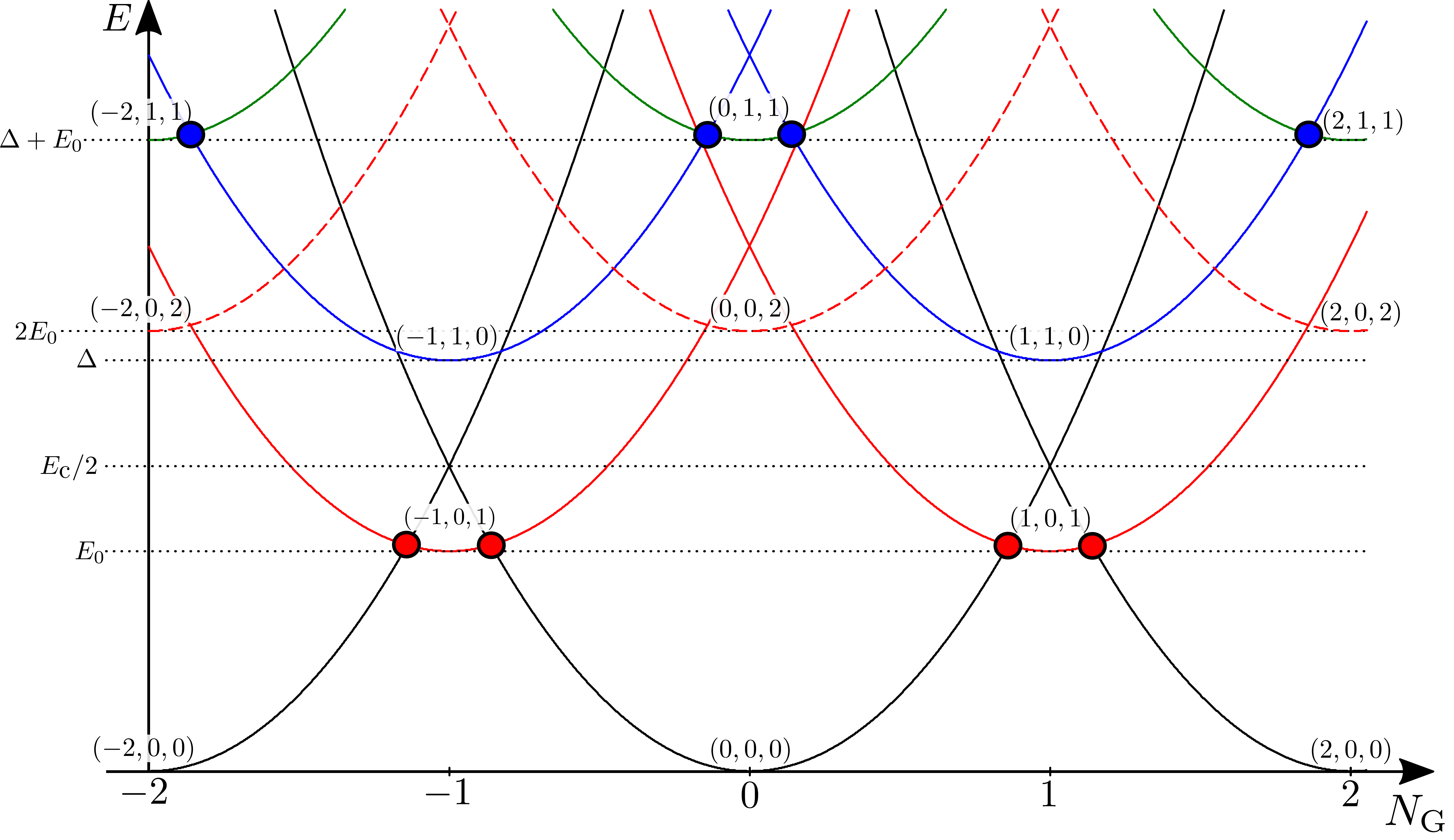}
  \caption{Spectrum resulting from Eq.~(\ref{eq.spectrum}) for different states $(N,N_\Delta,N_0)$ as a function of the gate-induced charge $N_{\text{G}}$. Red circles mark degeneracies between different charge states $(N_{\text{even}},0,0)$ and $(N_{\text{odd}},0,1)$, in the ground state branch. Blue circles mark degeneracies between different charge states $(N_{\text{even}},1,1)$ and $(N_{\text{odd}},1,0)$, in a branch of excited (poisoned) states. The plot is made to scale with the parameters given in the main text: $E_{\text{C}}=210~\mu\text{eV}$, $\Delta=140~\mu\text{eV}$, and $E_{0}=75~\mu\text{eV}$.}
  \label{fig.spectrum}
\end{figure}

The low-energy part of the resulting spectrum is plotted in Fig.~\ref{fig.spectrum} as a function of the gate-induced charge $N_{\text{G}}$, where we assumed $E_0 < E_{\rm C}/2 < \Delta < 2E_0$.
Each parabola corresponds to a state $(N,N_\Delta,N_0)$, the labels are included in the figure (note that $N$ is to be understood modulo an arbitrary number of Cooper pairs on the hybrid island).
For even $N$, the ground state is a pure BCS condensate without any excitations in the subgap state or in the BCS continuum (black curves in the plot).
The first excited state for even $N$ is a state with two excitations in the subgap state, having an excitation energy of $2E_0$ (red dashed curves).
The second excited state (solid green) has one excitation in the subgap state and one in the BCS continuum, resulting in an excitation energy of $\Delta+E_0$.
For odd $N$ we included the ground state (solid red), which has one excitation in the subgap state (energy $E_0$), and the first excited state (solid blue), where the excitation is in the BCS continuum instead (energy $\Delta$).
The next excited state has two excitations in the subgap state and one in the BCS continuum (energy $\Delta + 2 E_0$). This state is not shown in Fig.~\ref{fig.spectrum}, but it is included in our simulations.

\begin{figure}[hbt]
  \centering
  \subfloat[][$(N_{\text{even}},0,0)$]{
    \includegraphics[width=0.25\textwidth]{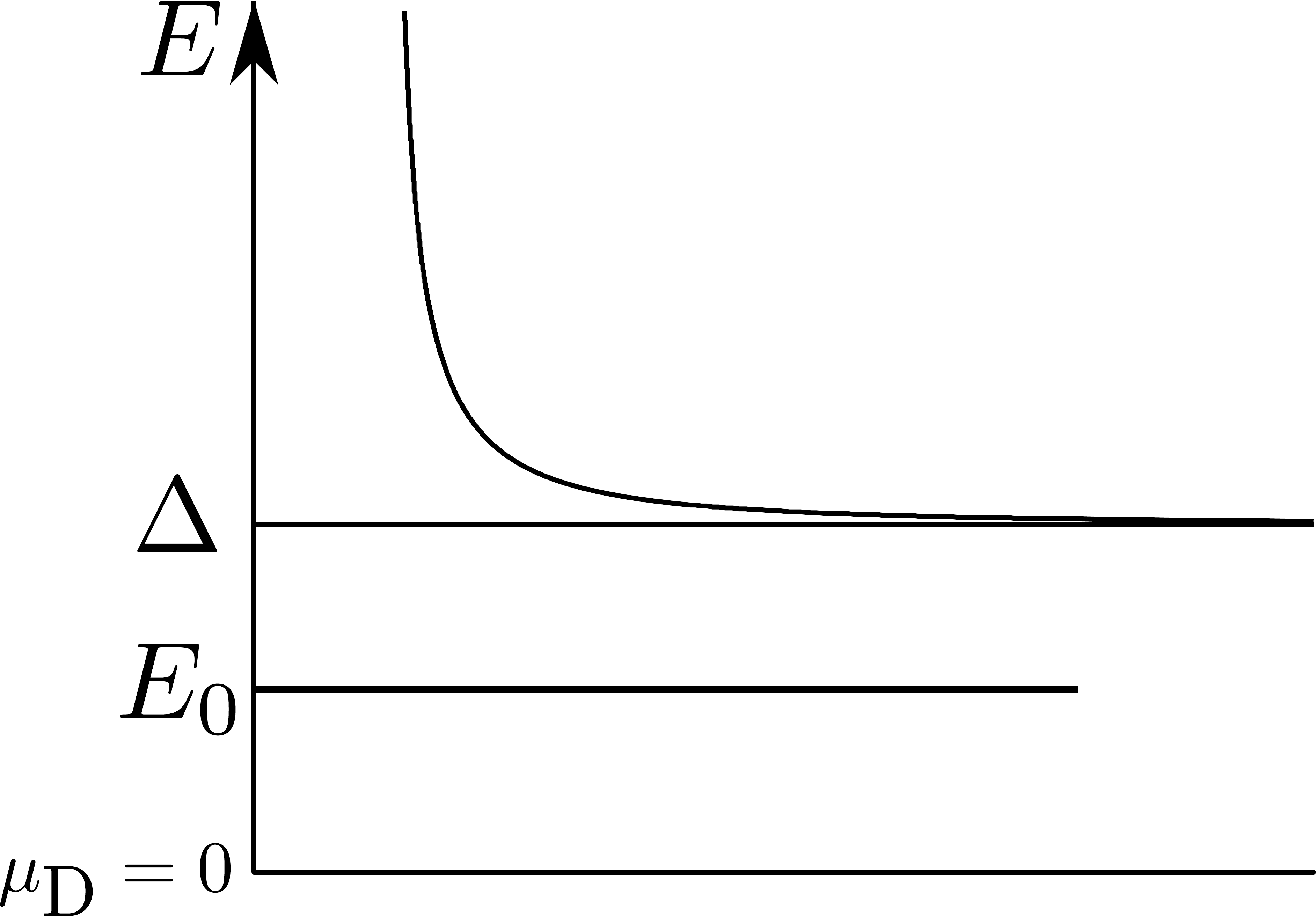}
%    \caption{A gull}
%    \label{fig:gull}
  }
  ~
  \subfloat[][$(N_{\text{even}},0,2)$]{
    \includegraphics[width=0.25\textwidth]{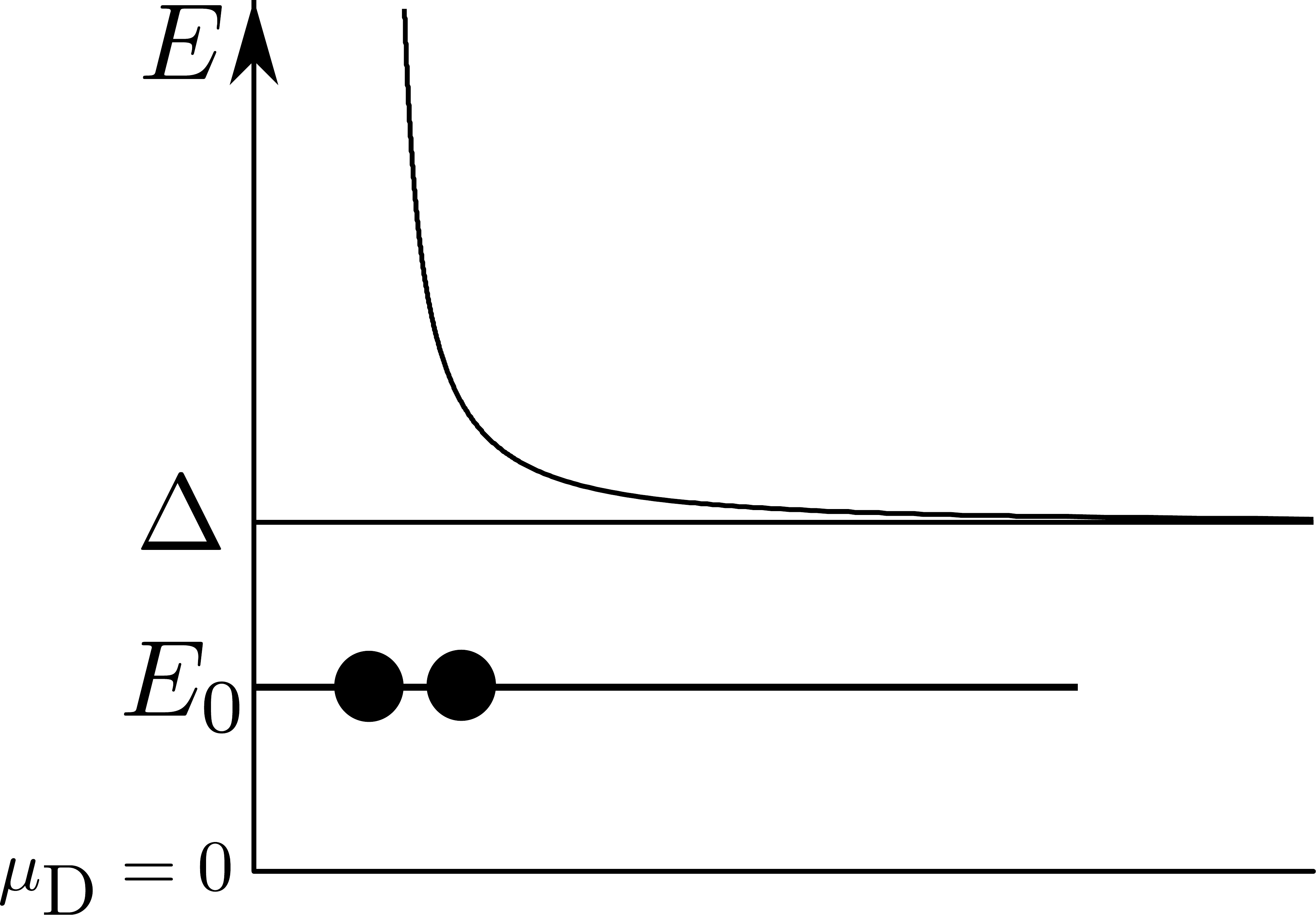}
%    \caption{A gull}
%    \label{fig:gull2}
  }
  ~
  \subfloat[][$(N_{\text{even}},1,1)$]{
    \includegraphics[width=0.25\textwidth]{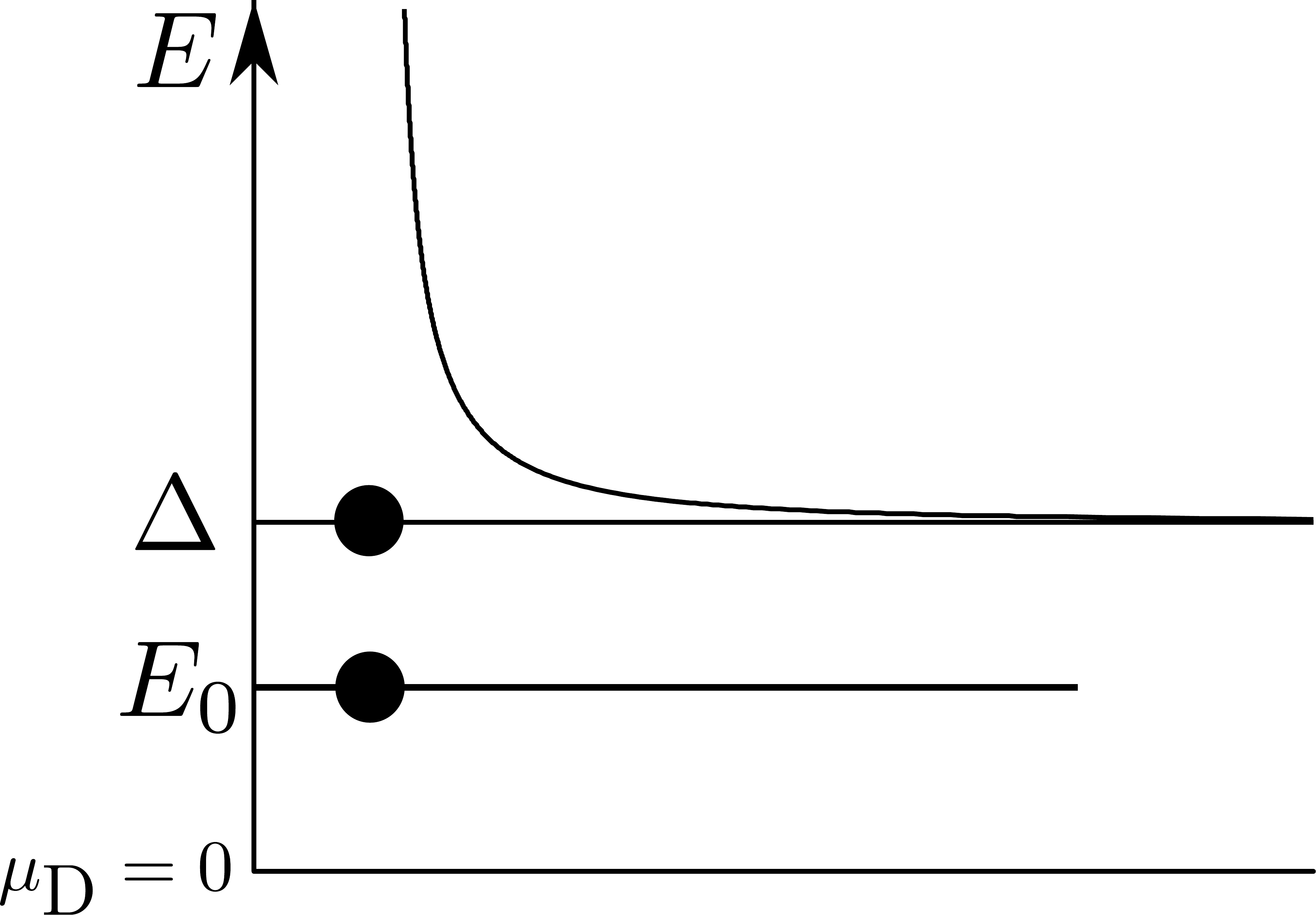}
%    \caption{A gull}
%    \label{fig:gull4}
  }
  
  \subfloat[][$(N_{\text{odd}},0,1)$]{
    \includegraphics[width=0.25\textwidth]{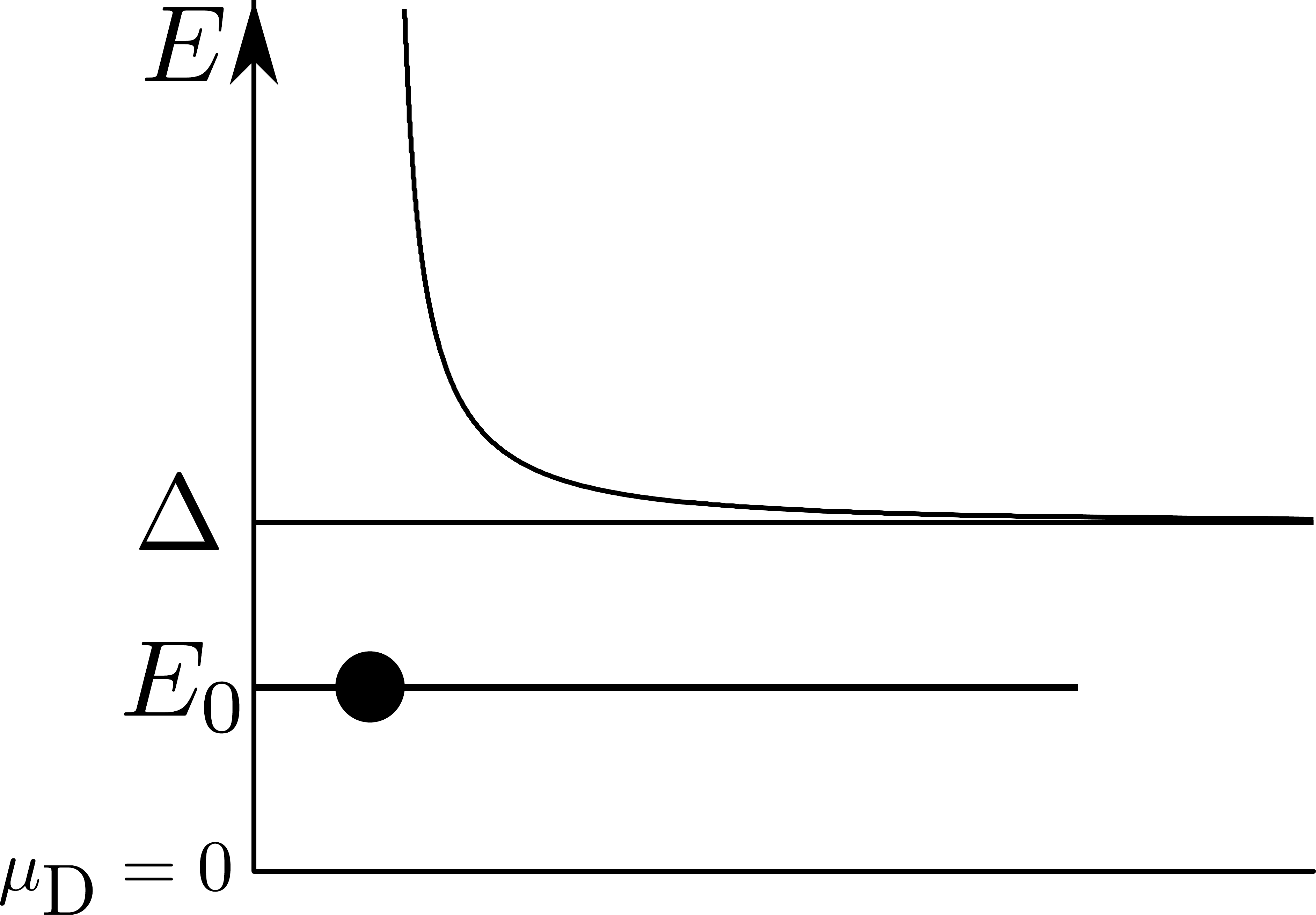}
%    \caption{A gull}
%    \label{fig:gull1}
  }
  ~
  \subfloat[][$(N_{\text{odd}},1,0)$]{
    \includegraphics[width=0.25\textwidth]{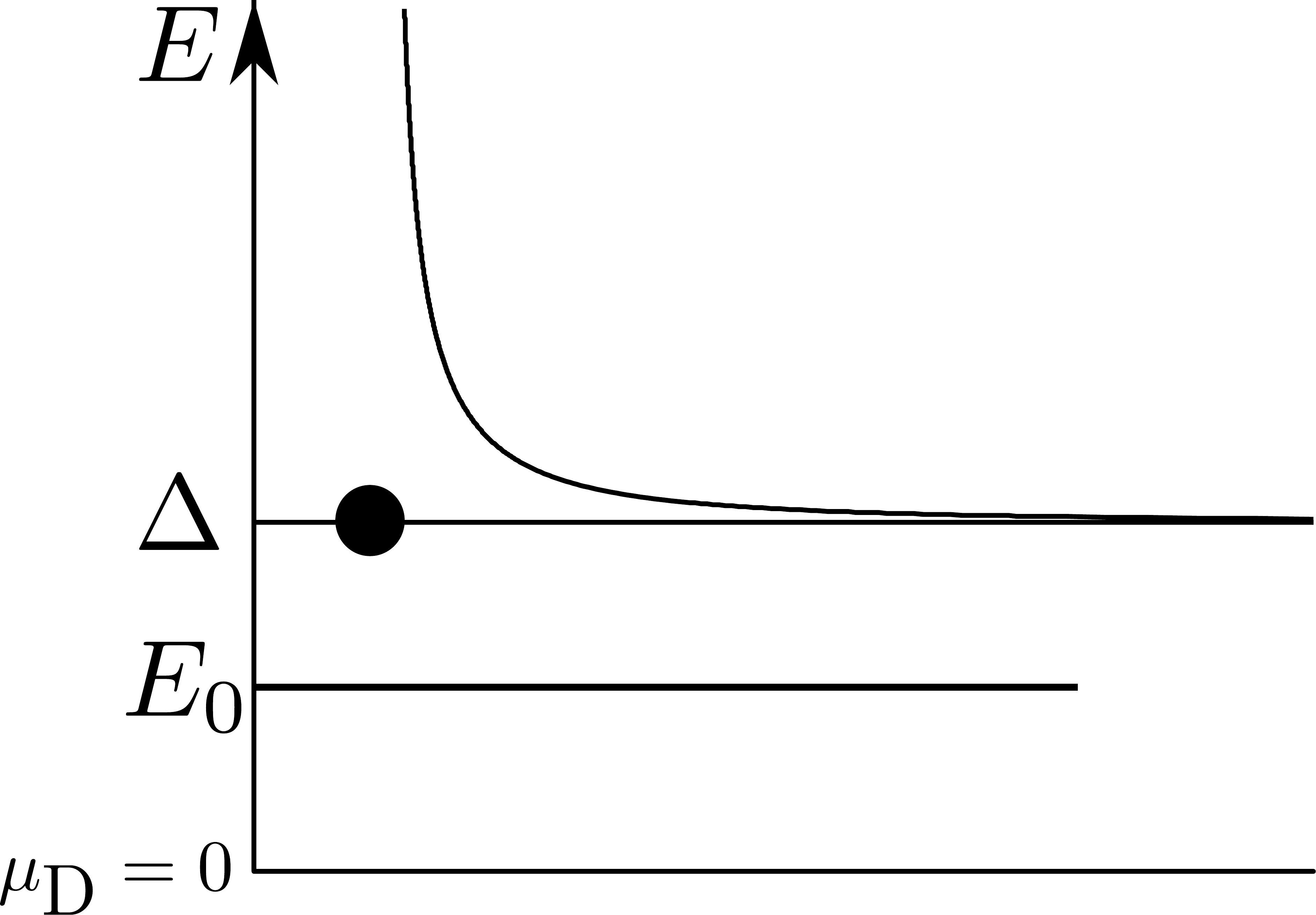}
%    \caption{A gull}
%    \label{fig:gull3}
  }
  \caption{Schematic density of states of the island. (a--c) Assuming even $N$, the  three lowest-energy states are (a) without any excitations, (b) a state with a doubly occupied subgap state, and (c) a state with a singly occupied subgap state and one quasiparticle excitation. (d,e) With odd $N$ the lowest-energy state (d) has a singly occupied subgap mode and the first excited state (e) has one quasiparticle excitation.}\label{fig.states}
\end{figure}

In Fig.~\ref{fig.states} we show a schematic of the density of states of the hybrid island for the five different types of charge states included in Fig.~\ref{fig.spectrum}.

\subsection{Transport}

We assume that transport is dominated by incoherent sequential tunneling processes, and we describe the dynamics of the island with a Pauli master equation.
We thus calculate the equilibrium probability distribution function $P_{(N,N_\Delta,N_0)}$ by solving the master equation
\begin{align}
  \frac{d}{dt} P_{(N,N_\Delta,N_0)} {} & {} = \sum_{N',N_\Delta',N_0'}\left\{ \Gamma_{(N,N_\Delta,N_0)\leftarrow (N',N_\Delta',N_0')}P_{(N',N_\Delta',N_0')} - \Gamma_{(N',N_\Delta',N_0')\leftarrow (N,N_\Delta,N_0)}P_{(N,N_\Delta,N_0)}\right\} \nonumber\\
  {} & {} = 0,
  \label{eq.pauli_master}
\end{align}
together with the normalization condition
\begin{align}
  \sum_{N,N_\Delta,N_0} P_{(N,N_\Delta,N_0)} = 1,
\end{align}
where $\Gamma_{(N',N'_\Delta,N'_0)\leftarrow (N,N_\Delta,N_0)}$ is the incoherent transition rate from the charge state $(N,N_\Delta,N_0)$ to $(N',N'_\Delta,N'_0)$.

Rates with $N\neq N'$ involve tunneling of charges to and from the leads, and these rates thus include contributions from both leads.
Only focusing on (lowest-order) single-particle tunneling, this means that we can write
\begin{align}
  \Gamma_{(N\pm 1,N'_\Delta,N'_0)\leftarrow (N,N_\Delta,N_0)} = \sum_{\alpha\in\{L,R\}}\Gamma_{(N\pm 1,N'_\Delta,N'_0)\leftarrow (N,N_\Delta,N_0)}^{(\alpha)},
\end{align}
where the superscript $\alpha$ indicates which lead the charge is tunneling to or from.
The current resulting from sequential tunneling is then obtained as
\begin{align}
  I = (-e)\sum_{N,N_\Delta, N_\Delta'} \left\{ \Gamma^{(L)}_{(N+1,N_\Delta',N_0')\leftarrow (N,N_\Delta,N_0)}-\Gamma^{(L)}_{(N-1,N_\Delta',N_0')\leftarrow (N,N_\Delta,N_0)}\right\}P_{(N,N_\Delta,N_0)}.
\end{align}

In order to calculate the current explicitly, the only task left is to find all relevant transition rates $\Gamma_{(N',N'_\Delta,N'_0)\leftarrow (N,N_\Delta,N_0)}$.
In our model we include (i) tunneling of charges to and from the leads resulting in the creation or annihilation of a quasiparticle excitation in the BCS continuum, $\Gamma^{(L,R)}_{(N\pm 1,N_\Delta \pm 1,N_0)\leftarrow (N,N_\Delta,N_0)}$, (ii) tunneling to and from the leads combined with a change of the occupation of the subgap mode $\Gamma^{(L,R)}_{(N\pm 1,N_\Delta,N_0')\leftarrow (N,N_\Delta,N_0)}$, and (iii) we add internal relaxation processes on the island $\Gamma_{(N,N_\Delta',N_0')\leftarrow (N,N_\Delta,N_0)}$.
In the following subsections we will discuss all transitions we included in more detail.

\subsubsection{Tunneling into and out of the BCS continuum}

We calculate all tunneling rates using Fermi's golden rule,
\begin{equation}
\Gamma_{\beta\leftarrow \alpha} = 2\pi \sum_{f_\beta, i_\alpha}|\langle f_\beta | H_{\text{T}} | i_\alpha\rangle |^2 W_{i_\alpha} \delta(E_{f_\beta}-E_{i_\alpha}),\label{eq.fermis_rule}
\end{equation}
where $\alpha = (N,N_\Delta,N_0)$ and $\beta = (N',N_\Delta',N_0')$ and we set $\hbar = 1$.
The sums are over all the configurations of internal degrees of freedom $i_\alpha$ that correspond to the initial state $\alpha$, each weighted by the appropriate thermal distribution function $W_{i\alpha}$, and all the configurations of internal degrees of freedom $f_\beta$ of that yield $\beta$.
The supplementary material of Ref.~\citesupp{Higginbotham2015} contains a detailed derivation of these rates.
Here, we will only present the results.

The rates for adding an electron to or removing one from the island by creating or annihilating a quasiparticle excitation in the BCS continuum are found to be
\begin{align}
\Gamma^{(\alpha)}_{(N+1,N_\Delta+\eta,N_0)\leftarrow (N,N_\Delta,N_0)} &= \gamma_\alpha\int_\Delta^\infty d E \left\{ 
\frac{E}{\sqrt{E^2-\Delta^2}}A_{\text{in},\eta}(E)\right\},\\
\Gamma^{(\alpha)}_{(N-1,N_\Delta + \eta,N_0)\leftarrow (N,N_\Delta,N_0)} &= \gamma_\alpha\int_\Delta^\infty d E \left\{ 
\frac{E}{\sqrt{E^2-\Delta^2}}A_{\text{out},\eta}(E)\right\},
\end{align}
where $\eta=+1$($-1$) applies to $N_\Delta=0$($1$), the coupling strength $\gamma_\alpha$ is related to the normal conductance $g_{\text{Al}} = 2\pi (e^2/h) \gamma_\alpha$ and
\begin{align}
A_{\text{in},\eta}(E) &= n_F(E_{\text{el}}^{(N)}+E-\mu_\alpha)[1-f_\eta(E)] + n_F(E_{\text{el}}^{(N)}-E-\mu_\alpha)f_\eta(E),\label{eq.A_in}\\
A_{\text{out},\eta}(E)  &= [1-n_F(E_{\text{el}}^{(N)}+E-\mu_\alpha)]f_\eta(E) + [1-n_F(E_{\text{el}}^{(N)}-E-\mu_\alpha)][1-f_\eta(E)]\label{eq.A_out},
\end{align}
where $n_F$ is the Fermi distribution function and $E_{\text{el}}^{(N)} = E_{\text{el}}(N+1)-E_{\text{el}}(N)$ the difference in electrostatic energy between states with $N+1$ and $N$ charges on the island.
The even/odd parity effect of the above-gap quasiparticles on the free energy is included in the distribution functions\cite{Tuominen1992,Lafarge1993}
\begin{equation}
  f_{\eta}(E) = \frac{1}{e^{\beta(E+\eta\,\delta F_\text{BCS})} + 1}.
\end{equation}
$\beta = 1/k_{\rm B}T$ is the inverse temperature, and for low temperatures $\beta\Delta\gg1$ the free energy difference $\delta F_\text{BCS}$ can be approximated as
\begin{align}
  \delta F_{\text{BCS}} &\approx -\frac{1}{\beta}\ln\tanh[2N_{\text{eff}}K_1(\beta\Delta)],
\end{align}
where $K_\nu(z)$ is the modified Bessel function of the second kind and $N_{\text{eff}}=\rho_D \Delta$ is the effective number of  quasiparticle states involved,
with $\rho_D$ the normal density of states at the Fermi level of the island including spin.

The first(last) term in Eq.~(\ref{eq.A_in}) corresponds to adding an electron to the island by creating(annihilating) a quasiparticle excitation on the island. In Eq.~(\ref{eq.A_out}), the first(last) term corresponds to removing an electron from the island by annihilating(creating) a quasiparticle excitation.

\subsubsection{Tunneling into and out of the subgap state}

Using similar methods we can calculate the rates of tunneling into and out of the subgap state.
The resulting transition rates are
\begin{align}
\Gamma^{(\alpha)}_{(N-1,N_\Delta,0)\leftarrow (N,N_\Delta,\sigma)} &= \tfrac{1}{2}\gamma^{\text{sub}}_\alpha [1-n_F(E^c_{N}-E^c_{N-1}+E_{0\sigma}-\mu_\alpha)],\\
\Gamma^{(\alpha)}_{(N+1,N_\Delta,0)\leftarrow (N,N_\Delta,\sigma)} &= \tfrac{1}{2}\gamma^{\text{sub}}_\alpha n_F(E^c_{N+1}-E^c_{N}-E_{0\sigma}-\mu_\alpha),\\
\Gamma^{(\alpha)}_{(N+1,N_\Delta,2)\leftarrow (N,N_\Delta,\sigma)} &= \tfrac{1}{2}\gamma^{\text{sub}}_\alpha n_F(E^c_{N+1}-E^c_N+E_{0\bar{\sigma}}-\mu_\alpha),\\
\Gamma^{(\alpha)}_{(N-1,N_\Delta,2)\leftarrow (N,N_\Delta,\sigma)} &= \tfrac{1}{2}\gamma^{\text{sub}}_\alpha [1-n_F(E^c_{N}-E^c_{N-1}-E_{0\bar{\sigma}}-\mu_\alpha)],\\
\Gamma^{(\alpha)}_{(N+1,N_\Delta,\sigma)\leftarrow (N,N_\Delta,0)} &= \tfrac{1}{2}\gamma^{\text{sub}}_\alpha n_F(E^c_{N+1}-E^c_{N}+E_{0\sigma}-\mu_\alpha),\\
\Gamma^{(\alpha)}_{(N+1,N_\Delta,\sigma)\leftarrow (N,N_\Delta,2)} &= \tfrac{1}{2}\gamma^{\text{sub}}_\alpha n_F(E^c_{N+1}-E^c_{N}-E_{0\bar{\sigma}}-\mu_\alpha),\\
\Gamma^{(\alpha)}_{(N-1,N_\Delta,\sigma)\leftarrow (N,N_\Delta,2)} &= \tfrac{1}{2}\gamma^{\text{sub}}_\alpha [1-n_F(E^c_{N}-E^c_{N-1}+E_{0\bar{\sigma}}-\mu_\alpha)],\\
\Gamma^{(\alpha)}_{(N-1,N_\Delta,\sigma)\leftarrow (N,N_\Delta,0)} &= \tfrac{1}{2}\gamma^{\text{sub}}_\alpha [1-n_F(E^c_{N}-E^c_{N-1}-E_{0\sigma}-\mu_\alpha)],
\end{align}
where $\sigma \in\{\uparrow,\downarrow\}$ and we have assumed that $|u_0|^2 = |v_0|^2 = 1/2$. The estimated coupling strength to the subgap state $\gamma^{\text{sub}}_\alpha$ is determined from fitting a zero-bias conductance peak to the functional form of a Breit-Wigner resonance with unequal tunnel barriers, see Supplement of Ref. \citesupp{Jorgensen2007}.

\subsubsection{Internal relaxation}

The negative differential conductance observed in Fig.~2(a) in the main text is associated with current blocking due to the occupation of a state in the BCS continuum.
These states are relatively weakly coupled to the leads, yielding a very slow BCS quasiparticle escape rate.\cite{Higginbotham2015,Hekking1993,Hergenrother1994}
However, the blocking quasiparticle can escape the island through a different process than directly tunneling out to a lead:
It can first relax into the subgap state, which is much more strongly coupled to the leads and thus facilitates fast subsequent tunneling out of a charge.
This relaxation process corresponds to the transition rates
\begin{equation}
  \Gamma_{(N,0,\sigma)\leftarrow(N,1,0)} = \Gamma_{(N,0,2)\leftarrow(N,1,\sigma)} = \Gamma_{(N,0,0)\leftarrow(N,1,\sigma)}
  = \Gamma_{(N,0,\sigma)\leftarrow(N,1,2)} \equiv \Gamma_{\text{relax}}.
\end{equation}
In terms of the schematics shown in Fig.~\ref{fig.states}: The first rate describes a transition from an initial state as pictured in (e) to a final state (d), the second rate corresponds to going from (c) to (b), and the third rate from (c) to (a).
The last rate describes transitions from a state with three excitations (two in the subgap and one in the continuum) to (d).
For simplicity we assume the same, energy-independent relaxation rate for all these transitions.
The reverse thermal excitation rates are also included and read
\begin{align}
  \Gamma_{(N,1,0)\leftarrow(N,0,\sigma)} &= \Gamma_{\text{relax}}e^{-\beta(\Delta - E_{0\sigma})},\\
  \Gamma_{(N,1,\sigma)\leftarrow(N,0,2)} &= \Gamma_{\text{relax}}e^{-\beta(\Delta - E_{0\bar{\sigma}})},\\
  \Gamma_{(N,1,\sigma)\leftarrow(N,0,0)} &= \Gamma_{\text{relax}}e^{-\beta(\Delta + E_{0\sigma})},\\
  \Gamma_{(N,1,2)\leftarrow(N,0,\sigma)} &= \Gamma_{\text{relax}}e^{-\beta(\Delta + E_{0\bar{\sigma}})}.
\end{align}
The associated relaxation time was quantified as $\tau_{\text{relax}} = \Gamma_\text{relax}^{-1} = 0.1~\mu$s in Ref.~\citesupp{Higginbotham2015}, where an similar device geometry with Majorana island length $L = 310~ \mathrm{nm} $ was used, tuned to a regime with weaker coupling between the island and the leads.
These relaxation processes are internal processes of the island and therefore should not depend on the coupling to the leads.
We thus included these processes in our simulations using the same relaxation time as found in Ref.~\citesupp{Higginbotham2015}, i.e., we set $\Gamma_\text{relax} = 10$~MHz.
This rate is necessary to quantitatively reproduce the observed negative differential conductance in our experiment, and it also plays an important role in estimating the quasiparticle poisoning rate, as will become clear below.

We also include the similar process of Cooper pair recombination(breaking) from(to) the doubly-occupied subgap state,
\begin{align}
\Gamma_{(N,0,0)\leftarrow(N,0,2)} = \Gamma_{(N,1,0)\leftarrow(N,1,2)} &= \Gamma_{\text{relax}},\\
\Gamma_{(N,0,2)\leftarrow(N,0,0)} = \Gamma_{(N,1,2)\leftarrow(N,1,0)} &= \Gamma_{\text{relax}}e^{-\beta(E_{0\sigma}+E_{0\bar{\sigma}})}.
\end{align}

\subsubsection{Quasiparticle poisoning}

Finally, we include quasiparticle poisoning processes that excite the island from its ground state to an excited state with one extra quasiparticle excitation in the BCS continuum.

At low (or zero) bias voltage and low temperatures, the island is expected to be in its ground state, which is $(N_{\text{even}},0,0)$ or $(N_{\text{odd}},0,1)$, depending on $N_{\rm G}$ (see Fig.~\ref{fig.spectrum}).
Sequential tunneling of charges from the leads onto and out of the island should produce peaks in the conductance whenever the ground state changes it total charge number, i.e., whenever $(N_{\text{even}},0,0)$ and $(N_{\text{odd}},0,1)$ are degenerate (see the red circles in Fig.~\ref{fig.spectrum}).
These conductance peaks are indeed clearly present in the data.

The observation of a set of ``shadow'' peaks in the center of the resulting Coulomb diamonds [Fig.~2(a) in the main text] indicates that the island is not always in its ground state.
The gate voltages at which these shadow peaks occur suggest that the peaks reflect transport through the island while it has one extra quasiparticle excitation in its BCS continuum.
Indeed, the gate voltages where $(N_{\text{even}},1,1)$ and $(N_{\text{odd}},1,0)$ are degenerate (blue circles in Fig.~\ref{fig.spectrum}) seem to agree with the gate voltages where the shadow peaks occur in the data.
We thus assume that there are significant poisoning rates adding an extra quasiparticle excitation to the island,
\begin{equation}
  \Gamma^{(\alpha)}_{(N\pm1,1,N_0)\leftarrow(N,0,N_0)} \equiv \Gamma_{\text{qp}},
  \label{eq:pois}
\end{equation}
which we add to our model.
While the island is in the resulting excited state with $N_\Delta = 1$, current can flow through the subgap state, $(N_{\text{even}},1,1) \rightleftarrows (N_{\text{odd}},1,0)$, and the height of the resulting conductance peak thus scales with the ratio $\Gamma_\text{qp}/\Gamma_\text{relax}$

The most likely physical mechanism responsible for such poisoning is a non-equilibrium (high-energy) electron or hole in one of the leads tunneling onto the island [the label $\alpha$ in Eq.~(\ref{eq:pois}) indicates from which lead the quasiparticle originates], thereby creating a quasiparticle excitation somewhere in the BCS continuum.
This excitation will relax relatively quickly to the edge of the continuum, where it has an  energy $\Delta$ and from where it can only escape by relaxation into the subgap state or tunneling out of the island.
We assume that the poisoning rate $\Gamma_\text{qp}$ is constant over the energy range we consider.

Another mechanism that could bring the island to a similar poisoned state is Cooper pair breaking:
A high-energy photon from the environment breaks a Cooper pair on the island into two quasiparticle excitations in the continuum.
Subsequently, one of these quasiparticles might relax into the subgap state, yielding the composite process
\begin{equation}
(N_\text{even},0,0)\xrightarrow[]{\Gamma_{\text{cpb}}}(N_\text{even},2,0)\xrightarrow[]{\Gamma_{\text{relax}}}(N_\text{even},1,1).
\label{eq:cpbpois}
\end{equation}
which effectively brings us in the same branch of excited states as described above, with a rate of roughly $\Gamma_\text{qp,cpb} = 1/(\Gamma_\text{cpb}^{-1} + \Gamma_\text{relax}^{-1})$, and can thus contribute to the shadow peaks depending on the ratio $\Gamma_\text{qp,cpb}/\Gamma_\text{relax}$.
However, in previous experiments where the same device was less strongly coupled to the leads no shadow peaks were observed \cite{Albrecht2016}. As Cooper pair breaking is a device-internal process and independent of device-lead coupling, this indicates that $\Gamma_\text{qp,cpb}/\Gamma_\text{relax} \ll 1$ and that in the present experiment the dominant process is poisoning from the leads.
We thus do not include the poisoning process (\ref{eq:cpbpois}) in our model, but we keep in mind that this process could become dominant in the less strongly coupled regime.

A last poisoning process we can consider to include is a non-equilibrium particle in the leads entering the subgap state, i.e.~transitions like $(N,N_\Delta,N_0)\to(N\pm 1,N_\Delta,N_0\pm 1)$.
However, first of all, this requires the particle to have a very specific energy ($E_{0\uparrow}$ or $E_{0\downarrow}$, up to the broadening of the subgap levels), and is therefore much less likely than the processes described above (although the subgap state is more strongly coupled to the leads than the BCS continuum).
Secondly, such a process will not have any significant effect on the transport through the island:
A charge entering the subgap state will quickly escape to one of the leads again, due to the strong coupling, and thus only result in a very small quantitative change of the current close to the ground state charge degeneracies.

\subsection{Zeeman splitting of the subgap states}\label{sec:zeeman}
\begin{figure}
  \includegraphics[width=12cm]{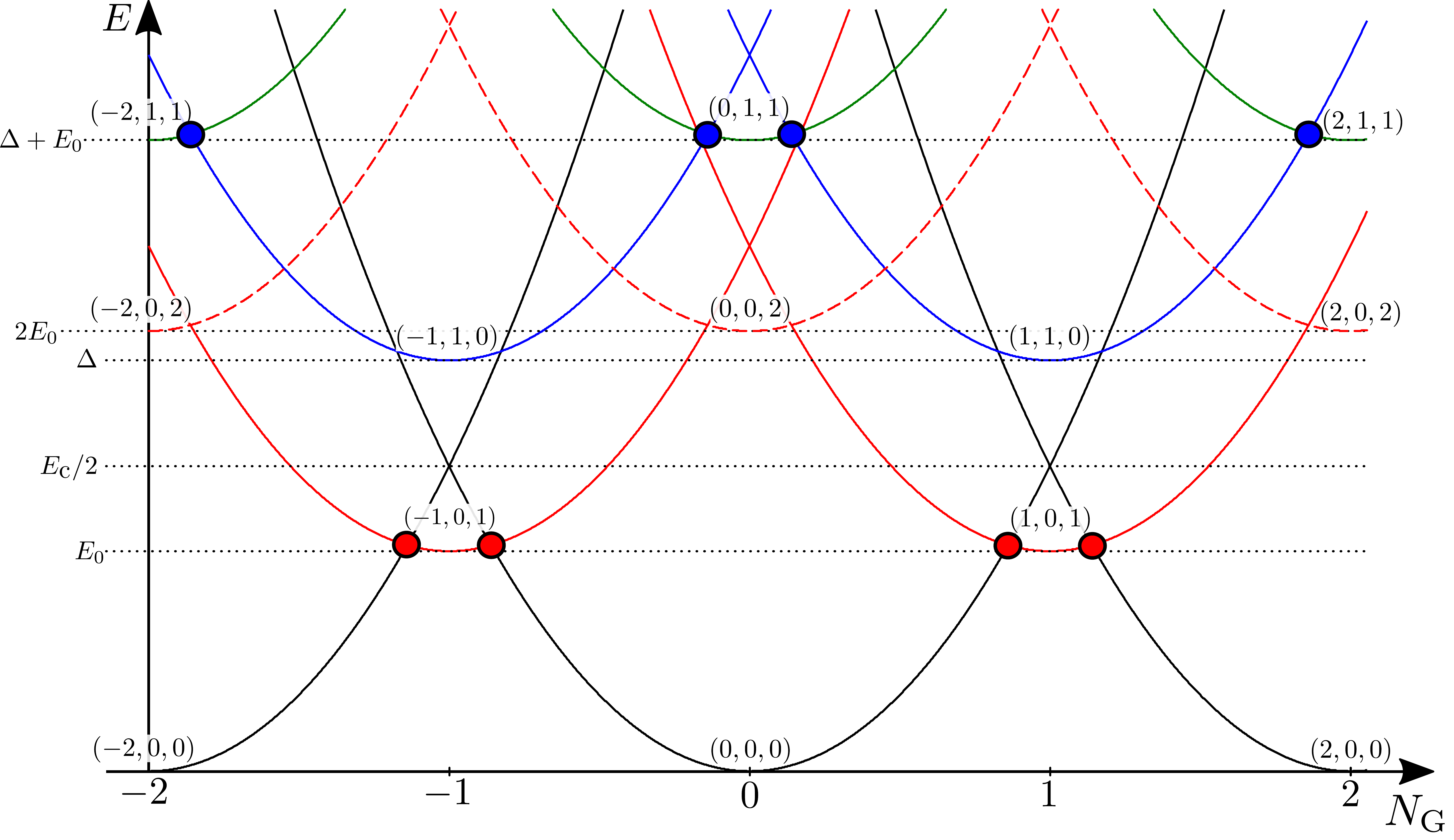}
  \caption{Spectrum as a function of the gate-induced charge $N_{\text{G}}$ resulting from Eq.~(\ref{eq.spectrum}) with $E_{0\sigma}$ as given in Eqs.~(\ref{eq.E_0_split_up}) and (\ref{eq.E_0_split_down}). Labels indicating the different states $(N,N_\Delta,N_0)$ are again included, and the plot is again made to scale with the parameters given in the main text: $E_{\text{C}}=210~\mu\text{eV}$, $\Delta=140~\mu\text{eV}$, and $E_{0}=75~\mu\text{eV}$. We now also included a Zeeman energy of $E_Z=25~\mu\text{eV}$.}
  \label{fig.spectrum2}
\end{figure}

Applying a magnetic field on the island will split the energy of the subgap states by a Zeeman splitting, which is linear for small fields.
Eventually, the state moving up in energy will merge with the (quasi-)continuum of BCS states, whereas the state moving down will evolve into a single low-energy mode (a Majorana mode in the limit of a long wire).
We model this effect by including a linear splitting of the two subgap states,
\begin{align}
  E_{0\uparrow}(B) &= \left\{
  \begin{array}{ll}E_{0} + E_{\rm Z} & \quad\text{for } 0 \leq E_{\rm Z} < \Delta - E_0, \\
  \Delta & \quad\text{for } E_{\rm Z} \geq \Delta - E_0, \end{array}\right. \label{eq.E_0_split_up} \\
  E_{0\downarrow}(B) &= \left\{
  \begin{array}{ll}E_{0} - E_{\rm Z} & \quad\text{for } 0 \leq E_{\rm Z} < E_0, \\
  0 & \quad\text{for } E_{\rm Z} \geq E_0, \end{array}\right. \label{eq.E_0_split_down}
\end{align}
where $E_{\rm Z} = \frac{1}{2}g\mu_{\rm B}B$ with $g$ the effective $g$-factor of the subgap state (taken positive here).
When the state with $\sigma = {\uparrow}$ reaches the superconducting continuum it stays at the energy $E_{0\uparrow}(B)=\Delta$ for larger $B$, and when the state with $\sigma = {\downarrow}$ reaches zero it stays at $E_{0\downarrow}(B)=0$ for larger $B$. We also include the effect of the decrease in coupling strength to the leads when the higher subgap state develops into a BCS continuum state in a phenomenological way, by making the coupling parameters spin- and field-dependent,
\begin{align}
  \gamma^\text{sub}_{\alpha,\uparrow} &= \gamma^\text{sub}_\alpha \left(1 - \frac{E_{0\uparrow}(B)-E_0}{\Delta-E_0}\right) + \gamma_\alpha\frac{E_{0\uparrow}(B)-E_0}{\Delta-E_0},\\
  \gamma^\text{sub}_{\alpha,\downarrow} &= \gamma^\text{sub}_\alpha.
\end{align}
For completeness, we show in Fig.~\ref{fig.spectrum2} again the spectrum as a function of gate-induced charge (similar as in Fig.~\ref{fig.spectrum}), but now including the effect of a finite Zeeman splitting.
We chose a field where both $E_{0\uparrow}$ and $E_{0\downarrow}$ are smaller than $E_{\rm C}/2$.

\end{document}